\documentclass[12pt]{article}

\usepackage[a4paper,text={16.3cm,22cm}]{geometry}
\usepackage{amsmath,amsfonts,braket,amssymb,bm,bbm,xcolor,graphicx}
\usepackage[small,labelfont=bf]{caption}
\usepackage{cite}
\usepackage[
    bookmarksnumbered=true,
    urlcolor=blue,
    linkbordercolor=red,
    citebordercolor=green,
    bookmarksopen=true
    ]{hyperref}

\allowdisplaybreaks
\numberwithin{equation}{section}

\newenvironment{Eqnarray}{\arraycolsep 0.14em\begin{eqnarray}}{\end{eqnarray}}
\def\beqa{\begin{Eqnarray}}
\def\eeqa{\end{Eqnarray}}
\newcommand{\no}{\nonumber}

\newcommand{\beq}{\begin{equation}}
\newcommand{\eeq}{\end{equation}}
\newcommand{\bea}{\begin{eqnarray}}
\newcommand{\eea}{\end{eqnarray}}

\title{Lessons from LHCb and Belle II measurements of \\[8pt]
$B\to J/\psi \pi$ and $B\to J/\psi K$ decays}
  
\begin{document}

\begin{titlepage}

\makeatletter
\vskip0.8cm
\begin{center}
{\Large \bf\boldmath \@title}
\end{center}
\makeatother

\vspace{0.5cm}
\begin{center}
Zoltan Ligeti,$^{a,b}$ Yosef Nir,$^c$ and Roy Schein$^d$ \\[6mm]
\textsl{${}^a$\,Theory Group, Lawrence Berkeley National Laboratory, Berkeley, 94720, USA\\[0.3cm]
\mbox{${}^b$\,Leinweber Institute for Theoretical Physics, University of California, Berkeley, 94720, USA}\\[0.3cm]
${}^c$\,Department of Particle Physics and Astrophysics, Weizmann Institute of Science\\ 
Rehovot 7610001, Israel\\ [0.3cm]
${}^d$\,Raymond and Beverly Sackler School of Physics and Astronomy, Tel Aviv 69978, Israel
}
\end{center}

\vspace{0.6cm}
\pdfbookmark[1]{Abstract}{abstract}
\begin{abstract}
The LHCb collaboration has recently measured the CP asymmetry in $B^+\to J/\psi\pi^+$ decay, while the Belle~II collaboration has recently measured the CP asymmetries in $B^0\to J/\psi\pi^0$ decay. Within the Standard Model, and using flavor-$SU(3)$ relations including first-order breaking corrections, these measurements lead to new predictions with regard to CP violation in $B^+\to J/\psi K^+$ and $B_s\to J/\psi\overline{K}{}^0$ decays, the difference between $S_{\psi K_S}$ and $\sin2\beta$, and the rate and CP asymmetries in $B_s\to J/\psi\pi^0$ decay.
\end{abstract}

\vfill\noindent\rule{0.4\columnwidth}{0.4pt}\\
\hspace*{2ex} {\small \textit{E-mail:} 
\href{mailto:ligeti@berkeley.edu}{ligeti@berkeley.edu}, 
\href{mailto:yosef.nir@weizmann.ac.il}{yosef.nir@weizmann.ac.il},
\href{mailto:schein1109@gmail.com}{schein1109@gmail.com}
}

\end{titlepage}

{\hypersetup{hidelinks}
\pdfbookmark[1]{Contents}{ToC}
\setcounter{tocdepth}{2}
\tableofcontents}
\vspace{4mm}

\section{Experimental results}
The LHCb collaboration has recently achieved the first evidence for direct CP violation in the $B^\pm\to J/\psi\pi^\pm$ decays~\cite{LHCb:2024exp}:
\beq\label{eq:bctopsipicexp}
\Delta {\cal A}^{CP}\equiv{\cal A}^{CP}(B^+\to J/\psi\pi^+)-{\cal A}^{CP}(B^+\to J/\psi K^+)=(+1.29\pm0.49\pm0.08)\times10^{-2}.
\eeq
The $B^+\to J/\psi \pi^+$ decay is one of six $B_q\to J/\psi P$ decays ($q=u,d,s$; $P=\pi,K$) related by $SU(3)$-flavor symmetry~\cite{Ligeti:2015yma,Jung:2012mp}. In principle, there are 16 relevant observables related to these six decay modes: six branching ratios (${\cal B}$), a CP asymmetry in each of the two charged $B$ decays (${\cal A}$), and two CP asymmetries, reflecting direct and indirect CP violation ($C$ and $S$), in each of the four neutral $B$ decays. Additional recent measurements of some of these observables include the LHCb measurement of the branching ratio of $B^0\to J/\psi\pi^0$~\cite{LHCb:2024ier}:
\beq\label{eq:bntopsipinexp}
{\cal B}(B^0\to J/\psi\pi^0)=(1.670\pm0.077\pm0.069\pm0.095)\times10^{-5},
\eeq
the Belle~II measurement of the CP asymmetries in $B^0\to J/\psi\pi^0$~\cite{Belle-II:2024hqw}:
\beqa
C^d_{\psi\pi}&=&+0.13\pm0.12\pm0.03,\no\\
S^d_{\psi\pi}&=&-0.88\pm0.17\pm0.03,\no\\
{\cal B}(B^0\to J/\psi\pi^0)&=&(2.00\pm0.12\pm0.09)\times10^{-5},
\eeqa
and the Belle upper bound on the branching ratio of $B_s\to J/\psi \pi^0$~\cite{Belle:2023tdz}:
\beq
{\cal B}(B_s\to J/\psi\pi^0)<1.21\times10^{-5}.
\eeq
For earlier measurements of CP asymmetries in the relevant decay modes, see Refs.~\cite{LHCb:2016ehk,BaBar:2008kfx,Belle:2018nxw,LHCb:2015brj,LHCb:2017joy}.
The current experimental status is summarized in Table~\ref{tab:expdata}.
Thus, of the sixteen observables, eight are measured (in the sense that they are established to be different from zero): five branching ratios and three CP asymmetries.

\begin{table}[!t]
	\caption{Current data. Taken from \cite{ParticleDataGroup:2024cfk,HeavyFlavorAveragingGroupHFLAV:2024ctg}, with updates from~\cite{LHCb:2024exp,LHCb:2024ier,Belle-II:2024hqw,Belle:2023tdz}. For ${\cal B}(B_d\to J/\psi\pi^0)$, we average the PDG value~\cite{ParticleDataGroup:2024cfk} with the recent LHCb~\cite{LHCb:2024ier} and Belle~II~\cite{Belle-II:2024hqw} results. For $S^d_{\psi\pi^0}$ and $C^d_{\psi\pi^0}$, we average over the HFLAV value~\cite{HeavyFlavorAveragingGroupHFLAV:2024ctg} and the recent Belle~II measurement~\cite{Belle-II:2024hqw}. For ${\cal A}_{\psi\pi^+}$ we interpret the LHCb result~\cite{LHCb:2024exp} with our Eq.~(\ref{eq:apiak}). } 
	\label{tab:expdata}
	\begin{center}
		\begin{tabular}{c|ccc} \hline\hline
			\rule{0pt}{1.2em}%
			Process & ${\cal B}$ & ${\cal A}/C$ & $S$ \cr
			\hline \hline
			$B^+\to J/\psi\pi^+$ & $(3.92\pm0.09)\times10^{-5}$ & $(+1.23\pm0.47)\times10^{-2}$ & $-$ \\
			$B^+\to J/\psi K^+$ & $(1.020\pm0.019)\times10^{-3}$ & $(+1.8\pm3.0)\times10^{-3}$ & $-$ \\
                         $B_d\to J/\psi\pi^0$ & $(1.74\pm0.07)\times10^{-5}$ & $(+8.5\pm8.5)\times10^{-2}$ & $-0.87\pm0.11$ \\
                         $B_d\to J/\psi K^0$ & $(8.91\pm0.21)\times10^{-4}$ & $(+0.9\pm1.0)\times10^{-2}$ & $+0.708\pm0.012$ \\
                         $B_s\to J/\psi\pi^0$ & $<1.21\times10^{-5}$ & NA & NA \\
                         $B_s\to J/\psi K_S$ & $(1.92\pm0.14)\times10^{-5}$ & $-0.28\pm0.41\pm0.08$ & $-0.08\pm0.40\pm0.08$ \\
			\hline\hline
		\end{tabular}
	\end{center}
\end{table}   

Our aim is to explore the implications of the new measurements for yet-unmeasured observables in Table~\ref{tab:expdata}. Furthermore, we discuss the implications for extracting the difference between the measured CP asymmetry in the golden mode, $S^d_{\psi K_S}$, and the value of the CP violating CKM parameter $\sin2\beta$~\cite{Ligeti:2015yma}. (For previous studies of this difference, see Refs.~\cite{Ciuchini:2005mg,Li:2006vq,Faller:2008zc,Gronau:2008cc,DeBruyn:2014oga,Frings:2015eva,Barel:2020jvf}. For an early use of SU(3) relations to constrain the difference between a variety of CP asymmetries and $\sin2\beta$, see \cite{Grossman:2003qp}.)

In Section~\ref{sec:theory} we present our formalism, the flavor-$SU(3)$ relations among the amplitudes of the six decay modes under study, and the resulting expressions for decay rates and CP asymmetries. In Section~\ref{sec:predictions} we obtain relations among various observables and find numerical predictions, based on the recent measurements. In Section~\ref{sec:hoc} we discuss the significance of higher order corrections to the relation between $S^d_{\psi K_S}$ and $\sin2\beta$. We conclude in Section~\ref{sec:conclusions}.

\section{Theoretical considerations}
\label{sec:theory}
\subsection{SU(3) relations}
In presenting the $SU(3)$ relations among the six decay amplitudes, we follow the formalism of Ref.~\cite{Ligeti:2015yma}. Each of the decay amplitudes can be written as follows:
\beq
A=\lambda_c^q A_c+\lambda^q_u A_u,\ \ \ \lambda^q_i\equiv V_{ib}^*V_{iq},
\eeq
with $i=u,c$ and $q=d,s$. We expand the amplitudes to first order in the $SU(3)$ breaking spurion $\varepsilon\,{\rm diag}\{1,1,-2\}$, and to zeroth order in the isospin breaking spurion $\delta\,{\rm diag}\{1,-1,0\}$. This is justified by the size of these spurions:
\beq
\varepsilon\sim(f_K/f_\pi)-1\sim0.2,\ \ \ \delta\sim(m_d-m_u)/\Lambda_{\chi{\rm SB}}\lesssim0.01.
\eeq
Thus, we decompose the amplitudes as follows:
\beqa\label{eq:su3breaking}
A_c&=&A_c^{(0)}+\varepsilon A_c^{(1)}+\cdots,\no\\*
A_u&=&A_u^{(0)}+\varepsilon A_u^{(1)}+\cdots,
\eeqa
The following CKM ratios play an important role in our analysis:
\beqa
\lambda_u^s/\lambda_c^s&=&\bar\lambda^2(\rho+i\eta),\no\\
\lambda_u^d/\lambda_c^d&=&-(\rho+i\eta),
\eeqa
where $\bar\lambda$, $\rho$, and $\eta$ are the Wolfenstein parameters, with
\beq
\bar\lambda^2=-\frac{{\cal R}e(\lambda_u^s/\lambda_c^s)}{{\cal R}e(\lambda_u^d/\lambda_c^d)}=0.0533\pm0.0003.
\eeq
The global CKM fit in the SM yields $\rho\approx0.16\pm0.01$ and $\eta\approx0.35\pm0.01$~\cite{ParticleDataGroup:2024cfk}. Given the smallness of the CKM ratios 
\beq
R_u^s\equiv|\lambda_u^s/\lambda_c^s|\simeq0.019,\qquad
R_u^d\equiv|\lambda_u^d/\lambda_c^d|\sim0.39, 
\eeq
we also neglect the $\lambda_u^q\varepsilon A_u^{(1)}$ terms.

There are five $SU(3)$ relations among the $A_c^{(0)}$ terms, four among the $A_c^{(1)}$ terms and three among the $A_u^{(0)}$ terms. Thus, we are left with six independent complex $A_q^{(0,1)}$ amplitudes. We define them as follows:
\beqa
A_c&\equiv&A_c^{(0)}(B^+\to J/\psi\pi^+),\no\\
A_c^\pi&\equiv&A_c^{(1)}(B^+\to J/\psi\pi^+),\no\\
A_c^K&\equiv&A_c^{(1)}(B^+\to J/\psi K^+),\no\\
A_u^+&\equiv&A_u^{(0)}(B^+\to J/\psi\pi^+),\no\\
A_u^0&\equiv&A_u^{(0)}(B^0\to J/\psi K^0),\no\\
A_u^\pi&\equiv&\sqrt{2}A_u^{(0)}(B^0\to J/\psi\pi^0).
\eeqa
In terms of these amplitudes, the $A(B_q\to J/\psi P)$ decay amplitudes are given by
\beqa
A(B^+\to J/\psi \pi^+)&=&\lambda_c^d(A_c+\varepsilon A_c^\pi)+\lambda_u^d A_u^+,\no\\
A(B_d\to J/\psi \pi^0)&=&-(1/\sqrt2)\lambda_c^d(A_c+\varepsilon A_c^\pi)+(1/\sqrt2)\lambda_u^d A_u^\pi,\no\\
A(B_s\to J/\psi \overline{K}{}^0)&=&\lambda_c^d[A_c-\varepsilon(A_c^\pi+A_c^K)]+\lambda_u^d A_u^0,\no\\
A(B^+\to J/\psi K^+)&=&\lambda_c^s(A_c+\varepsilon A_c^K)+\lambda_u^s A_u^+,\no\\
A(B_d\to J/\psi K^0)&=&\lambda_c^s(A_c+\varepsilon A_c^K)+\lambda_u^s A_u^0,\no\\
A(B_s\to J/\psi \pi^0)&=&(1/\sqrt2)\lambda_u^s(A_u^\pi+A_u^0).
\eeqa
%

\subsection{Decay rates}
To ${\cal O}(\varepsilon R_u, R_u^2)$, the rates (with straightforward modifications for $B_{d,s}\to J/\psi\pi^0$) are given by
\beqa\label{eq:ratesacau}
\Gamma(B\to f)&=&|\lambda_c^{q}|^2 |A_c|^2\left\{1+2\varepsilon{\cal R}e(A_c^{(1)}/A_c)+2{\cal R}e[(\lambda_u^q/\lambda_c^q)(A_u^{(0)}/A_c)]\right\},\no\\*
\Gamma(\overline{B}\to \overline{f})&=&|\lambda_c^{q}|^2 |A_c|^2\left\{1+2\varepsilon{\cal R}e(A_c^{(1)}/A_c)+2{\cal R}e[(\lambda_u^q/\lambda_c^q)(A_u^{(0)*}/A_c^*)]\right\}.
\eeqa
The CP-averaged decay rate is then given by
\beq
\hat\Gamma(B\to f)=|\lambda_c^{q}|^2 |A_c|^2\left[1+2\varepsilon{\cal R}e(A_c^{(1)}/A_c)+2{\cal R}e(\lambda_u^q/\lambda_c^q){\cal R}e(A_u^{(0)}/A_c)\right].
\eeq
Then,
\beqa\label{eq:hatgammasm}
\hat\Gamma(B^+\to J/\psi \pi^+)&=&|\lambda_c^d|^2|A_c|^2\left[1+2\varepsilon{\cal R}e( A_c^\pi/A_c)+2{\cal R}e(\lambda_u^d/\lambda_c^d){\cal R}e(A_u^+/A_c)\right],\no\\
\hat\Gamma(B_d\to J/\psi \pi^0)&=&(1/2)|\lambda_c^d|^2|A_c|^2\left[1+2\varepsilon{\cal R}e( A_c^\pi/A_c)-2{\cal R}e(\lambda_u^d/\lambda_c^d){\cal R}e(A_u^\pi/A_c)\right],\no\\
\hat\Gamma(B_s\to J/\psi \overline{K}{}^0)&=&|\lambda_c^d|^2|A_c|^2\left[1-2\varepsilon{\cal R}e(( A_c^\pi+A_c^K)/A_c)+2{\cal R}e(\lambda_u^d/\lambda_c^d){\cal R}e(A_u^0/A_c)\right],\no\\
\hat\Gamma(B^+\to J/\psi K^+)&=&|\lambda_c^s|^2|A_c|^2\left[1+2\varepsilon{\cal R}e( A_c^K/A_c)+2{\cal R}e(\lambda_u^s/\lambda_c^s){\cal R}e(A_u^+/A_c)\right],\no\\
\hat\Gamma(B_d\to J/\psi K^0)&=&|\lambda_c^s|^2|A_c|^2\left[1+2\varepsilon{\cal R}e( A_c^K/A_c)+2{\cal R}e(\lambda_u^s/\lambda_c^s){\cal R}e(A_u^0/A_c)\right],\no\\
\hat\Gamma(B_s\to J/\psi \pi^0)&=&(1/2)|\lambda_u^s|^2|A_u^\pi+A_u^0|^2.
\eeqa
%

\subsection{Direct CP violation}
The CP asymmetries in charged meson decays are defined as follows:
\beq
{\cal A}_f=\frac{{\cal B}(B^-\to f^-)-{\cal B}(B^+\to f^+)}{{\cal B}(B^-\to f^-)+{\cal B}(B^+\to f^+)}.
\eeq
Using Eq.~(\ref{eq:ratesacau}), we obtain
\beqa
{\cal A}_f&=&\frac{2|\lambda_c^q|^2|A_c|^2}{\hat\Gamma(B\to f)}{\cal I}m\left(\frac{\lambda_u^q}{\lambda_c^q}\right){\cal I}m\left(\frac{A_u^{(0)}}{A_c}\right) \no\\
&\simeq&2{\cal I}m(\lambda_u^q/\lambda_c^q){\cal I}m(A_u^{(0)}/A_c).
\eeqa
Specifically, we have
\beqa\label{eq:acharged}
{\cal A}_{\psi\pi^+}&=&2{\cal I}m(\lambda_u^d/\lambda_c^d){\cal I}m(A_u^+/A_c),\no\\
{\cal A}_{\psi K^+}&=&2{\cal I}m(\lambda_u^s/\lambda_c^s){\cal I}m(A_u^+/A_c).
\eeqa

The smallness of the CP asymmetries in semileptonic $B_d$ and $B_s$ decays implies that the neutral meson mixing parameters for both systems fulfill, to an excellent approximation,
\beq
|q/p|=1.
\eeq
Consequently, the above equations are valid also for the CP asymmetries $C_f^d$ in the $B_d$ and $C_f^s$ in the $B_s$ decays:
\beqa\label{eq:cqf}
C^d_{\psi\pi^0}&=&2{\cal I}m(\lambda_u^d/\lambda_c^d){\cal I}m(A_u^\pi/A_c),\no\\*
C^d_{\psi K^0}&=&2{\cal I}m(\lambda_u^s/\lambda_c^s){\cal I}m(A_u^0/A_c),\no\\
C^s_{\psi \overline{K}{}^0}&=&2{\cal I}m(\lambda_u^d/\lambda_c^d){\cal I}m(A_u^0/A_c),\no\\*
C^s_{\psi\pi^0}&=&0.
\eeqa
%
 
\subsection{Indirect CP violation}
Defining
\beq
\lambda_f=(q/p)(\overline{A}_f/A_f),
\eeq
then, for the relevant $B_{d,s}$ decay modes,
\beq
S_f=\frac{2{\cal I}m(\lambda_f)}{1+|\lambda_f|^2}.
\eeq

For the four neutral decay modes, we have
\beqa\label{eq:lambdafcp}
\lambda^d_{\psi K_S}&=&-e^{-2i\beta}\left[1-2i(A_u^0/A_c){\cal I}m(\lambda_u^{s}/\lambda_c^{s})\right],\no\\
\lambda^d_{\psi \pi^0}&=&-e^{-2i\beta}\left[1-2i(A_u^\pi/A_c){\cal I}m(\lambda_u^{d}/\lambda_c^{d})\right],\no\\
\lambda^s_{\psi K_S}&=&-e^{+2i\beta_s}\left[1-2i(A_u^0/A_c){\cal I}m(\lambda_u^{d}/\lambda_c^{d})\right],\no\\
\lambda^s_{\psi \pi^0}&=&-e^{+2i(\beta_s-\gamma)},
\eeqa
where
\beqa
\beta&\equiv&{\rm arg}\left(-\frac{V_{cd}V_{cb}^*}{V_{td}V_{tb}^*}\right),\quad
\beta_s\equiv{\rm arg}\left(-\frac{V_{ts}V_{tb}^*}{V_{cs}V_{cb}^*}\right),\no\\
\gamma&\equiv&{\rm arg}\left(-\frac{V_{ud}V_{ub}^*}{V_{cd}V_{cb}^*}\right)
={\rm arg}\left(\frac{V_{us}V_{ub}^*}{V_{cs}V_{cb}^*}\right)+{\cal O}(\lambda^4),
\eeqa
We remind the reader that in Eq.~(\ref{eq:lambdafcp}) we neglect terms of ${\cal O}(\varepsilon R_u^q)$ and ${\cal O}[(R_u^q)^2]$, which for $q=d$ are not very small. These are discussed below, in Section~\ref{sec:hoc}. 

With these approximations, we obtain for the $S_f$ asymmetries:
\beqa\label{eq:sbeta}
S^d_{\psi K_S}&=&+s_{2\beta}+2c_{2\beta}{\cal I}m(\lambda_u^s/\lambda_c^s){\cal R}e(A_u^0/A_c),\no\\
S^d_{\psi \pi^0}&=&-s_{2\beta}-2c_{2\beta}{\cal I}m(\lambda_u^d/\lambda_c^d){\cal R}e(A_u^\pi/A_c),\no\\
S^s_{\psi K_S}&=&-s_{2\beta_s}+2c_{2\beta_s}{\cal I}m(\lambda_u^d/\lambda_c^d){\cal R}e(A_u^0/A_c),\no\\*
S^s_{\psi \pi^0}&=&-\sin(2\gamma-2\beta_s).
\eeqa
%

\section{Predictions}
\label{sec:predictions}
\subsection{Decay rates}
To zeroth order in $\varepsilon$ and $R_u^q$, five of the six rates are equal up to an overall factor of $|\lambda_c^q|^2$ which distinguishes between the $b\to s$ and $b\to d$ transitions. Two of these approximate equalities hold also to first order in $\varepsilon$:
\beqa\label{eq:rudthe}
R^{ud}_{K^+K^0}&\equiv&\frac{\hat\Gamma(B^+\to J/\psi K^+)}{\hat\Gamma(B_d\to J/\psi K^0)}=1+2{\cal R}e\frac{\lambda_u^s}{\lambda_c^s}
{\cal R}e\frac{A_u^+-A_u^0}{A_c},\no\\
R^{ud}_{\pi^+\pi^0}&\equiv&\frac{\hat\Gamma(B^+\to J/\psi \pi^+)}{2\hat\Gamma(B_d\to J/\psi\pi^0)}=1+2{\cal R}e\frac{\lambda_u^d}{\lambda_c^d}
{\cal R}e\frac{A_u^++A_u^\pi}{A_c}.
\eeqa
The experimental values for these ratios are
\beqa\label{eq:rudexp}
R^{ud}_{K^+K^0}&=&1.06\pm0.03,\no\\
R^{ud}_{\pi^+\pi^0}&=&1.04\pm0.05.
\eeqa
Two more ratios of interest are the following:
\beqa\label{eq:ruursdthe}
R^{uu}_{\pi^+K^+}&\equiv&\left|\frac{V_{cs}}{V_{cd}}\right|^2\frac{\hat\Gamma(B^+\to J/\psi \pi^+)}{\hat\Gamma(B^+\to J/\psi K^+)}
=1+2\varepsilon{\cal R}e\frac{A_c^\pi-A_c^K}{A_c}+2{\cal R}e\left(\frac{\lambda_u^d}{\lambda_c^d}-\frac{\lambda_u^s}{\lambda_c^s}\right)
{\cal R}e\frac{A_u^\pi}{A_c},\no\\
R^{sd}_{\overline{K}{}^0K^0}&\equiv&\left|\frac{V_{cs}}{V_{cd}}\right|^2\frac{\hat\Gamma(B_s\to J/\psi \overline{K}{}^0)}{\hat\Gamma(B_d\to J/\psi K^0)}
=1-2\varepsilon{\cal R}e\frac{A_c^\pi+2A_c^K}{A_c}+2{\cal R}e\left(\frac{\lambda_u^d}{\lambda_c^d}-\frac{\lambda_u^s}{\lambda_c^s}\right)
{\cal R}e\frac{A_u^0}{A_c}.\qquad
\eeqa
The experimental values for these ratios are
\beqa\label{eq:ruursdexp}
R^{uu}_{\pi^+K^+}&=&0.72\pm0.02,\no\\
R^{sd}_{\overline{K}{}^0K^0}&=&0.81\pm0.06.
\eeqa
Given that $\varepsilon\sim0.2$ and 
\beq
{\cal R}e\left(\frac{\lambda_u^d}{\lambda_c^d}-\frac{\lambda_u^s}{\lambda_c^s}\right)\approx-0.16\pm0.01,
\eeq
the deviations from unity of ${\cal O}(0.2-0.3)$ for $R^{uu}_{\pi^+K^+}$ and $R^{sd}_{\overline{K}{}^0K^0}$ are as expected.

Using Eq.~(\ref{eq:hatgammasm}), we have
\beq
\frac{\hat\Gamma(B_s\to J/\psi\pi^0)}{\hat\Gamma(B\to J/\psi K)}=
\frac12\left|\frac{\lambda_u^s}{\lambda_c^s}\right|^2\left|\frac{A_u^\pi+A_u^0}{A_c}\right|^2.
\eeq
Ref.~\cite{Ligeti:2015yma} suggested a way to put a lower bound on ${\cal B}(B_s\to J/\psi\pi^0)$ by using the following two pairs of decays, where isospin relates the leading amplitudes in each pair~\cite{Jung:2015yma}:
\beqa\label{eq:Deltah}
\Delta_K&\equiv&\frac{\hat\Gamma(B_d\to J/\psi K^0)-\hat\Gamma(B^+\to J/\psi K^+)}{\hat\Gamma(B_d\to J/\psi K^0)+\hat\Gamma(B^+\to J/\psi K^+)},\no\\
\Delta_\pi&\equiv&\frac{2\hat\Gamma(B_d\to J/\psi\pi^0)-\hat\Gamma(B^+\to J/\psi\pi^+)}{2\hat\Gamma(B_d\to J/\psi\pi^0)+\hat\Gamma(B^+\to J/\psi\pi^+)}.
\eeqa
(There is a relation between the $\Delta$'s defined here and the respective $R$-ratios defined in Eq.~(\ref{eq:rudthe}): $R=1-2\Delta$.)
From Eq.~(\ref{eq:hatgammasm}), we obtain:
\beqa\label{eq:deltakdeltapi}
\Delta_K&=&-{\cal R}e(\lambda_u^s/\lambda_c^s){\cal R}e[(A_u^+-A_u^0)/A_c],\no\\
\Delta_\pi&=&-{\cal R}e(\lambda_u^d/\lambda_c^d){\cal R}e[(A_u^++A_u^\pi)/A_c],
\eeqa
leading to
\beq
\Delta_K+\bar\lambda^2\Delta_\pi
={\cal R}e\left(\frac{\lambda_u^s}{\lambda_c^s}\right){\cal R}e\left(\frac{A_u^0+A_u^\pi}{A_c}\right).
\eeq
Then,
\beqa\label{eq:lowerboundBspsipithe}
\frac{\hat\Gamma(B_s\to J/\psi\pi^0)}{\hat\Gamma(B\to J/\psi K)}
&\geq&\frac{(\Delta_K+\bar\lambda^2\Delta_\pi)^2}{2\cos^2\gamma},\no\\
{\cal B}(B_s\to J/\psi\pi^0)&\geq&\frac{\tau_{B_s}}{\tau_{B^+}}
\frac{(\Delta_K+\bar\lambda^2\Delta_\pi)^2}{2\cos^2\gamma}\times{\cal B}(B^+\to J/\psi K^+).
\eeqa

Using the branching ratios from Table~\ref{tab:expdata} for an indirect determination of $\Delta_K$, as well as four recent direct measurements by Belle~\cite{BELLE:2019xld}, $\Delta_K=-0.002\pm0.006\pm0.014$, Belle~II~\cite{Belle-II:2022dbo}, $\Delta_K^{J/\psi(ee)}=-0.022\pm0.016\pm0.030$, and $\Delta_K^{J/\psi(\mu\mu)}=-0.006\pm0.015\pm0.030$, and LHCb~\cite{LHCb:2025jva}, $\Delta_K=-0.0135\pm0.0133$, we obtain:\footnote{The ${\cal B}(B\to J/\psi K)$ measurements in Ref.~\cite{BELLE:2019xld} are also included in the PDG averages, however, the direct measurement of $\Delta_K$ (called $A_I$ in~\cite{BELLE:2019xld}) has a smaller uncertainty, and their correlation is not specified.} 
\beq\label{eq:deltakexp}
\Delta_K=-0.015\pm0.008.
\eeq
We note that there is a subtlety in the extraction of $\Delta_K$ from the experimental measurements in the $B$ factories related to isospin violation in $\Upsilon(4S)\to B^+B^-$ vs.\ $B^0\overline{B}{}^0$. Ref.~\cite{Bernlochner:2023bad} studies this uncertainty and obtains an average over the $\Upsilon(4S)$ measurements of $\Delta_K=-0.002\pm0.017$. However, for the LHCb result, we expect negligible production asymmetry.

Using the branching ratios from Table~\ref{tab:expdata} to determine $\Delta_\pi$,  we obtain:
\beqa\label{eq:deltapiindirect}
\Delta_\pi&=&-0.021\pm0.023,
\eeqa
and hence,
\beq\label{eq:delklamdelpiexp}
\Delta_K+\bar\lambda^2\Delta_\pi=-0.016\pm0.008.
\eeq

Using the central values of $\gamma$~\cite{ParticleDataGroup:2024cfk}, $\gamma=(65.7\pm3.0)^\circ$ (so that $\cos^2\gamma=0.17\pm0.04$), and of Eq.~(\ref{eq:delklamdelpiexp}), we obtain:
\beqa\label{eq:lowerboundBspsipiexp}
\hat\Gamma(B_s\to J/\psi\pi^0)&\gtrsim&8.5\times10^{-4}\ \hat\Gamma(B\to J/\psi K),\no\\*
{\cal B}(B_s\to J/\psi\pi^0)&\gtrsim&9\times10^{-7},
\eeqa
where we used $\tau_{B_s}/\tau_{B^+}\simeq0.93$ and the central value of ${\cal B}(B^+\to J/\psi K^+)$ from Table~\ref{tab:expdata}. The lower bound is a factor of 13 below the current experimental upper bound. Note, however, that given the uncertainty of $\Delta_K+\bar\lambda^2\Delta_\pi$ (or using the $\Delta_K$ value of Ref.~\cite{Bernlochner:2023bad}), Eq.~(\ref{eq:lowerboundBspsipithe}) currently provides no lower bound at~$2\sigma$.

\subsection{Direct CP violation}
Taking into account that
\beq
-{\cal I}m(\lambda_u^d\lambda_c^{d*})=+{\cal I}m(\lambda_u^s\lambda_c^{s*})=J_{\rm CKM},
\eeq
Eq.~(\ref{eq:acharged}) leads to
\beq
\frac{{\cal A}_{\psi K^+}}{{\cal A}_{\psi\pi^+}}=-\left|\frac{V_{cd}}{V_{cs}}\right|^2=-0.053.
\eeq
This leads to the following interpretation concerning Eq.~(\ref{eq:bctopsipicexp}):
\beqa\label{eq:apiak}
\Delta{\cal A}^{CP}&\equiv&{\cal A}_{\psi\pi^+}-{\cal A}_{\psi K^+}\simeq{\cal A}_{\psi\pi^+}(1+|V_{cd}/V_{cs}|^2)\no\\
{\cal A}_{\psi\pi^+}&\simeq&(+1.23\pm0.47)\times10^{-2},\no\\
{\cal A}_{\psi K^+}&\simeq&(-6.5\pm2.5)\times10^{-4}.
\eeqa

Additional relation concerning direct CP violation is the following:
\beq
\frac{C^d_{\psi K^0}}{C^s_{\psi\overline{K}{}^0}}=-\left|\frac{V_{cd}}{V_{cs}}\right|^2=-0.053,
\eeq
leading to
\beqa
C^d_{\psi K^0}=(+0.9\pm1.0)\times10^{-2}&~\Longrightarrow~&C^s_{\psi\overline{K}{}^0}=-0.17\pm0.19,\no\\
C^s_{\psi\overline{K}{}^0}=-0.28\pm0.42&~\Longrightarrow~&C^d_{\psi K^0}=(+1.5\pm2.2)\times10^{-2}.
\eeqa
We learn that the predictions on the RHS are consistent with the current experimental ranges on the LHS. Moreover, the prediction for $C^s_{\psi\overline{K}{}^0}$ is more precise than current measurements.

Within our approximations, we further predict
\beq\label{eq:cspsipi}
C^s_{\psi\pi^0}=0.
\eeq
One should be cautious however about this prediction. In our analysis, we neglected isospin breaking and, in particular, terms of the form $\lambda^s_c\,\delta A_c^{(2)}$. If these terms are not much smaller than the $\lambda_u^s A_u^{(0)}$ terms that we did take into account, then $C^s_{\psi\pi^0}$ is not necessarily $\ll1$.  

\subsection{Indirect CP violation}
Eq. (\ref{eq:sbeta}) leads to
\beq\label{eq:spsikss2b}
S^d_{\psi K_S}-s_{2\beta}=-\left|\frac{V_{cd}}{V_{cs}}\right|^2\frac{c_{2\beta}}{c_{2\beta_s}}\times(S^s_{\psi K_S}+s_{2\beta_s}).
\eeq
We neglect here the second order corrections. Their significance is discussed in Section~\ref{sec:hoc}.
With current values, $S^s_{\psi K_S}=-0.08\pm0.41$, $s_{2\beta_s}=+0.040\pm0.016$ (and thus $c_{2\beta_s}\simeq1$), $c_{2\beta}\approx0.7$ and $|V_{cd}/V_{cs}|^2\simeq0.053$, we obtain
\beq\label{eq:spsikss2bnum}
S^d_{\psi K_S}-s_{2\beta}=-0.037\times(-0.04\pm0.41)=+0.001\pm0.015.
\eeq
An experimental effort to improve the measurement of $S^s_{\psi K_S}$ is called for. Reducing the uncertainties from ${\cal O}(0.4)$ to below ${\cal O}(0.3)$ will determine $S^d_{\psi K_S} - \sin2\beta$ to better than one percent. 

Another way to determine this difference, which turns out to be less precise at present, was suggested in Ref~\cite{Ligeti:2015yma}. Eqs.~(\ref{eq:sbeta}) can be rewritten as
\beqa\label{eq:sbeta2}
S^d_{\psi K_S}&=&s_{2\beta}+2c_{2\beta}(J_{\rm CKM}/|\lambda_c^s|^2){\cal R}e(A_u^0/A_c),\no\\
S^d_{\psi \pi^0}&=&-s_{2\beta}-2c_{2\beta}(J_{\rm CKM}/|\lambda_c^d|^2){\cal R}e(A_u^\pi/A_c).
\eeqa
Using $|V_{cd}/V_{cs}|^2\simeq\bar\lambda^2$, we find
\beqa\label{eq:s2bLR}
S^d_{\psi K_S}-\bar\lambda^2S^d_{\psi\pi^0}&=&(1+\bar\lambda^2)\sin2\beta+2c_{2\beta}(J_{\rm CKM}/|\lambda_c^s|^2){\cal R}e[(A_u^0+A_u^\pi)/A_c]\no\\*
&=&(1+\bar\lambda^2)\sin2\beta+2c_{2\beta}t_\gamma(\Delta_K+\bar\lambda^2\Delta_\pi).
\eeqa
Putting the experimental ranges in Eq.~(\ref{eq:s2bLR}) (we use $c_{2\beta}=0.67$ and $\tan\gamma=2.2\pm0.3$):
\beqa\label{eq:s2bLRnum}
\sin2\beta&=&\frac{S^d_{\psi K_S}-\bar\lambda^2 S^d_{\psi\pi^0}-2c_{2\beta}t_\gamma(\Delta_K+\bar\lambda^2\Delta_\pi)}{1+\bar\lambda^2}\no\\
&=&\frac{0.708\pm0.012+0.046\pm0.006+0.050\pm0.029}{1.053}
 = +0.76\pm0.03.
\eeqa
The central value, which corresponds to $\sin2\beta-S^d_{\psi K_S}\sim0.05$ is surprisingly large, but it is less than $2\sigma$ away from $\sin2\beta=S^d_{\psi K_S}$. One should await more precise measurements of the various relevant observables, $\Delta_K$ in particular (also requiring improved determination of isospin violation in $\Upsilon(4S)\to B^+B^-$ vs.\ $B^0\overline B{}^0$~\cite{Bernlochner:2023bad}), to obtain a more precise extraction of $\sin2\beta$ using this relation.  

Using
\beq
S^s_{\psi \pi^0}=-\sin(2\gamma-2\beta_s)=-s_{2\gamma}c_{2\beta_s}+c_{2\gamma}s_{2\beta_s},
\eeq
we predict
\beq
S^s_{\psi\pi^0}=-0.78\pm0.07.
\eeq
Again, this prediction is based on the assumption that the $\lambda_c^s\,\delta A_c^{(2)}$ contribution can be neglected.

The relations in Eqs.~(\ref{eq:spsikss2b}) and (\ref{eq:s2bLR}) require largely independent experimental inputs.  While LHCb can measure all ingredients in Eq.~(\ref{eq:spsikss2b}), improving the measurement of $S^d_{\psi\pi^0}$ entering Eq.~(\ref{eq:s2bLR}) can probably only be done at Belle~II.

\section{Higher order corrections}
\label{sec:hoc}
In the relations that we use to extract $S^d_{\psi K_S}-s_{2\beta}$, our goal is to achieve high precision. It is then important to check that the higher-order corrections do not introduce uncertainties at the desired level of accuracy. 

We focus on the relation in Eq.~(\ref{eq:spsikss2b}). To second order in $\eta,\rho,\varepsilon$, we have
\beqa\label{eq:sdsshoc}
S^d_{\psi K_S}-s_{2\beta}&=&+2c_{2\beta}\bar\lambda^2\eta\left\{
{\cal R}e(A_u^0/A_c)\left[1-\varepsilon{\cal R}e(A_c^K/A_c)-\bar\lambda^2\rho{\cal R}e(A_u^0/A_c)\right]\right.\no\\
&&\left.+{\cal I}m(A_u^0/A_c)\left[\varepsilon{\cal I}m(A_c^K/A_c)+\bar\lambda^2\rho{\cal I}m(A_u^0/A_c)\right]\right\},\no\\
S^s_{\psi K_S}+s_{2\beta_s}&=&-2c_{2\beta_s}\eta\left\{
{\cal R}e(A_u^0/A_c)\left[1+\varepsilon{\cal R}e((A_c^\pi+A_c^K)/A_c)+\rho{\cal R}e(A_u^0/A_c)\right]\right.\no\\
&&\left.+{\cal I}m(A_u^0/A_c)\left[-\varepsilon{\cal I}m((A_c^\pi+A_c^K)/A_c)-\rho{\cal I}m(A_u^0/A_c)\right]\right\}.
\eeqa

The corrections to the leading order results divide into a multiplicative factor and an additive term. Interestingly, for the additive term, one can use Eqs.~(\ref{eq:cqf}) and (\ref{eq:sbeta}) at leading order to replace theoretical amplitude ratios with experimentally measurable observables:
\beqa
{\cal I}m(A_u^0/A_c)&=&-C^s_{\psi K_S}/(2\eta),\no\\
{\cal R}e(A_u^0/A_c)&=&-S^s_{\psi K_S}/(2c_{2\beta_s}\eta),
\eeqa
where in the second relation we neglected an $s_{2\beta_s}$ term compared to $S^s_{\psi K_S}$.
We then obtain:
\beqa
\frac{S^d_{\psi K_S}-s_{2\beta}}{S^s_{\psi K_S}+s_{2\beta_s}}&=&
-\bar\lambda^2\frac{c_{2\beta}}{c_{2\beta_s}}\times
\Bigg[1-\varepsilon{\cal R}e\frac{A_c^\pi+2A_c^K}{A_c}+\rho(1+\bar\lambda^2)\frac{S^s_{\psi K_S}}{2c_{2\beta_s}\eta}\no\\
&& + \frac{C^s_{\psi K_S}c_{2\beta_s}}{S^s_{\psi K_S}}\left(\varepsilon{\cal I}m\frac{A_c^\pi+2A_c^K}{A_c}-\rho(1+\bar\lambda^2)\frac{C^s_{\psi K_S}}{2\eta}\right)\Bigg].
\eeqa
Finally, examining Eq.~(\ref{eq:ruursdthe}), we can further replace the multiplicative correction with an experimentally measured observable:
\beqa\label{eq:bpsikhoc}
\frac{S^d_{\psi K_S}-s_{2\beta}}{S^s_{\psi K_S}+s_{2\beta_s}}=
-\bar\lambda^2\frac{c_{2\beta}}{c_{2\beta_s}}
\left[1+\frac{R^{sd}_{\overline{K}{}^0K^0}-1}2
+ \frac{C^s_{\psi K_S}c_{2\beta_s}}{S^s_{\psi K_S}}\left(\varepsilon{\cal I}m\frac{A_c^\pi+2A_c^K}{A_c}-\rho(1+\bar\lambda^2)\frac{C^s_{\psi K_S}}{2\eta}\right)\right]. \no\\
\eeqa
From Eq.~(\ref{eq:ruursdexp}), we learn that $(R^{sd}_{\overline{K}{}^0K^0}-1)/2\approx-0.10\pm0.03$, consistent with $SU(3)$-breaking of ${\cal O}(0.2)$. The remaining corrections, proportional to $\varepsilon$ and $\rho$,  are also expected to be of ${\cal O}(0.2)$. With future improved measurements of $S^s_{\psi K_S}$ and $C^s_{\psi K_S}$, one will be able to assess the significance of these unknown higher-order terms.

The relations (\ref{eq:spsikss2b}) and (\ref{eq:bpsikhoc}) call for an experimental effort to close in on $S^s_{\psi K_S}$ and $C^s_{\psi K_S}$. It will pave the way for what may potentially become the best determination of the difference between $S^d_{\psi K_S}$ and $\sin2\beta$ (especially if $|C^s_{\psi K_S}/S^s_{\psi K_S}|\lesssim1$). 

However, in this discussion, there is a missing ingredient: As presented in Eq.~(\ref{eq:su3breaking}), $SU(3)$ breaking applies also to the $A_u$ terms. We neglected these $A_u^{(1)}$ terms, as their contributions are proportional to $\lambda_u^q\varepsilon$. This procedure is justified for decay rates, 
where they generate corrections of order $R_u^q\varepsilon$, but not when discussing higher-order corrections to the CP asymmetries, where the leading contributions are of order $R_u^q$ and the $A_u^{(1)}$ terms generate corrections of order $\varepsilon$.

Let us first note that, for the corresponding $A_c^{(1)}$ terms, Eq.~(\ref{eq:ruursdexp}) gives $\varepsilon{\cal R}e\big[A_c^{(1)}(B_d\to J/\psi K^0)-A_c^{(1)}(B_s\to J/\psi \overline{K}{}^0)\big]\approx-0.10\pm0.03$. If the $A_u^{(1)}$ amplitudes follow a similar pattern, we should expect corrections of the same order.   

While there are no $SU(3)$ symmetry relations among the six $A_u^{(1)}$ amplitudes, dynamical arguments suggest, specifically for the $B_d\to J/\psi K_S$ and $B_s\to J/\psi K_S$ decays, that the difference between the respective $A_u^{(1)}$ terms may be small. Although in the heavy quark limit these decays formally factorize when the charm quark is treated as heavy~\cite{Beneke:2000ry}, it has long been recognized that phenomenologically it provides a poor approximation for the $b\to c\bar cq$ contribution, since the $c\bar c$ pair is nearly on-shell (see, e.g., \cite{Boos:2004xp,Frings:2015eva}). However, we expect that the $b\to u\bar uq$ contribution is better approximated by perturbative techniques. The $u\bar u$ pair (or the $J/\psi$ meson) is a $U$-spin singlet, so the difference between $A_u$ in these two decays comes from the $K_S$ being formed from a $d$-quark in the Hamiltonian and an $\bar s$-spectator, or an $s$-quark in the Hamiltonian and a $\bar d$-spectator. The kinematical differences are small, and in an explicit calculation of the $B\to K^0$ and $B_s\to \overline{K}{}^0$ form factors, their difference should trace the small strange-quark mass dependence of the $B_s$ meson's light-cone distribution amplitude~\cite{Feldmann:2023aml}.\footnote{Comparing lattice QCD results for $B\to K^0$ and $B_s\to \overline{K}{}^0$ is hindered by the fact that in most works either the spectator flavor or the flavor of the light quark in the weak current is varied, but not both. The comparison of $f_+$ and $f_0$ in earlier calculations~\cite{Bouchard:2013eph, Bouchard:2014ypa} supports a small difference at $q^2 = m_{J/\psi}^2$.}
 
If the difference between the $A_u^{(1)}$'s of the two decays can be neglected, then Eq.~(\ref{eq:bpsikhoc}) remains valid. If not, then there will be further corrections proportional to $\varepsilon\big[A_u^{(1)}(B_d\to J/\psi K^0)-A_u^{(1)}(B_s\to J/\psi \overline{K}{}^0\big]$.   

\section{Conclusions}
\label{sec:conclusions}
The six decay modes of $B_q\to J/\psi P$, with $q=u,d,s$ and $P=\pi,K$, obey various relations due to the $SU(3)$-flavor symmetry of QCD. Recently, a significant progress has been achieved by the LHCb and Belle~II experiments in measuring rates and CP asymmetries of some of these modes. We used these new experimental results to make predictions for yet unmeasured observables:
\begin{itemize}
\item Using
\beq\label{eq:akapi}
\frac{{\cal A}_{\psi K^+}}{{\cal A}_{\psi\pi^+}}=-\left|\frac{V_{cd}}{V_{cs}}\right|^2,
\eeq
we are able to extract the individual CP asymmetries from the LHCb measurement of $\Delta {\cal A}^{CP}$: ${\cal A}_{\psi\pi^+}\simeq(1.23\pm0.47)\times10^{-2}$ and ${\cal A}_{\psi K^+}\simeq(-6.5\pm2.5)\times10^{-4}$.
\item Using
\beq\label{eq:cdkcsk}
\frac{C^d_{\psi K^0}}{C^s_{\psi\overline{K}{}^0}}=-\left|\frac{V_{cd}}{V_{cs}}\right|^2,
\eeq
we narrow the allowed range for $C^s_{\psi\overline{K}{}^0}=-0.17\pm0.19$.
\item Using
\beq\label{eq:sdkssk}
S^d_{\psi K_S}-s_{2\beta}=-\left|\frac{V_{cd}}{V_{cs}}\right|^2\frac{c_{2\beta}}{c_{2\beta_s}}\times(S^s_{\psi K_S}+s_{2\beta_s}),
\eeq
we find $S^d_{\psi K_S}-\sin2\beta=+0.001\pm0.015$.
This constraint could improve substantially from an updated measurement of $S^s_{\psi K_S}$.
\item Using
\beq\label{eq:sdksdpi}
\sin2\beta=\frac{S^d_{\psi K_S}-\bar\lambda^2S^d_{\psi\pi^0}-2c_{2\beta}t_\gamma(\Delta_K+\bar\lambda^2\Delta_\pi)}{1+\bar\lambda^2},
\eeq
we obtain an estimate of the difference $\sin2\beta-S^d_{\psi K_S}=+0.05\pm0.03$.
\item Our predictions concerning the $B_s\to J/\psi\pi^0$ decay assume that the isospin violating $\lambda_c^s\delta A_c^{(2)}$ contribution can be neglected. Using 
\beq
\frac{{\cal B}(B_s\to J/\psi\pi^0)}{{\cal B}(B^+\to J/\psi K^+)}
\geq\frac{\tau_{B_s}}{\tau_{B^+}}\frac{(\Delta_K+\bar\lambda^2\Delta_\pi)^2}{2\cos^2\gamma},
\eeq
we obtain a lower bound, ${\cal B}(B_s\to J/\psi\pi^0)\gtrsim9\times10^{-7}$ by putting in central values for the measured observables. At $2\sigma$, however, there is as yet no bound. For the CP asymmetries, we predict $C^s_{\psi\pi^0}=0$, and $S^s_{\psi\pi^0}=-\sin(2\gamma-2\beta_s)=-0.78\pm0.07$.
\end{itemize}

The system of the six $B_q\to J/\psi P$ decays provides potentially 16 observables while, in the approximations that we make (neglecting terms of ${\cal O}(\varepsilon R_u,R_u^2,\delta)$), depending on 12 real parameters. The resulting four relations are presented in Eqs.~(\ref{eq:akapi}), (\ref{eq:cdkcsk}), (\ref{eq:sdkssk}) and (\ref{eq:sdksdpi}). Experiments are making significant progress in measuring the relevant observables, thus enabling tests of the SM, predictions for the yet unmeasured observables, and precision determination of the CKM parameters. Further improving the experimental precision of the eight measured observables, and adding to the list any of the eight yet-unmeasured ones, will significantly improve our ability to test the SM. 

While this paper was in completion, Ref.~\cite{DeBruyn:2025rhk} appeared, which studies related issues.

\section*{Acknowledgements}
We thank Yuval Grossman for useful discussions. We thank Dean Robinson for useful discussions and constructive criticism that led to significant improvements in our work.
YN is supported by a grant from the Minerva Foundation (with funding from the Federal Ministry for Education and Research). 
ZL is supported in part by the Office of High Energy Physics of the U.S.\ Department of Energy under contract DE-AC02-05CH11231. 



\end{document}